\documentclass[aps,prd,showpacs,twocolumn,a4paper]{revtex4-1}
\usepackage{graphicx}
\usepackage{amsmath}
\usepackage{multirow}
\usepackage{tabularx}
\usepackage{array}
\usepackage[usenames,dvipsnames]{xcolor}
\usepackage{hyperref}
\input{glyphtounicode}
\def\vermark{}

\begin{document}
\makeatletter
\newcommand{\ps@mytitle}{%
  \renewcommand{\@oddhead}{{\vermark}\hfil}%
}
\makeatother

\title{Directional dark matter by polar angle direct detection and application of columnar recombination}

\author{Jin Li}
\email{jinlee@ibs.re.kr}
\affiliation{Center for Underground Physics, Institute for Basic Science (IBS), Daejeon 305-811, Korea} 

\begin{abstract}
We report a systematic study on the directional sensitivity of a direct dark matter detector that detects the polar angle of a recoiling nucleus.  A weakly interacting massive particle (WIMP)-mass independent method is used to obtain the sensitivity of a general detector in an isothermal galactic dark matter halo.  By using two-dimensional distributions of energy and polar angle, a detector without head-tail information with 6.3 times the statistics is found to achieve the same performance level as a full three-dimensional tracking dark matter detector.   Optimum operation orientations are obtained for various experimental configurations, with detectors that are space- or Earth-fixed, have head-tail capability or not, and use energy information or not.  Earth-fixed detectors are found to have best sensitivity when the polar axis is oriented at a 45 degree angle from the Earth's pole.  With background contamination that mimics the WIMP signal's energy distribution, the performance is found to decrease at a rate less than the decrease of signal purity.  The WIMP-mass dependence of the performance of a detector with various energy thresholds that uses gaseous xenon as target material is reported.   We find that with a $5\times 10^{-46} \mathrm{cm}^2$ spin-independent WIMP-nucleon cross-section and a 30 GeV WIMP, a $770$ kg$\cdot$year's exposure with a polar detector of 10 keV threshold can make a three sigma discovery of directional WIMPs in the isothermal galactic dark matter halo.  For a columnar recombination detector, experimental considerations are discussed. 
\end{abstract}
\maketitle
\thispagestyle{mytitle}

\section{Introduction}
Many experiments have attempted to directly detect Weakly Interacting Massive Particle (WIMP) dark matter candidates via their elastic scattering on target nuclei~\cite{Agashe:2014kda}.  The LUX experiment has recently limited the spin-independent cross sections to be under $7.6\times 10^{-46}\ \mathrm{cm}^2$ for 33 GeV WIMP mass~\cite{Akerib:2013tjd}.  A method that has commonly been used is to measure signals associated with the deposited energy of the nuclear recoil.  However, background processes, such as those induced by neutrons, can mimic WIMP signals~\cite{Davis:2014cja}.  To overcome this problem, a ``smoking gun'' WIMP signal would be its unique directional event-rate dependence~\cite{Spergel:1987kx}.  When viewed from the Earth, the average WIMP velocity in galactic coordinates is from the Cygnus direction, while the directions of background sources are fixed in an Earth-based coordinate system.  This effect can be seen by the directions of the recoiling nuclei that are strongly correlated with the directions of the incoming WIMPs~\cite{Green:2006cb}, and produce a diurnal variation of rates due the Earth's rotation~\cite{Vergados:2006gw,Bandyopadhyay:2010zj}.  Prospects for a working directional detector have focused on low pressure gas time projection chambers, such as DRIFT-II~\cite{Alner:2005xp}, DMTPC~\cite{Battat:2014mka}, NEWAGE~\cite{Miuchi:2010hn}, MIMAC~\cite{Santos:2013hpa}, and the D$^3$ prototype~\cite{Vahsen:2014fba}, and emulsion techniques~\cite{Naka:2007zz}.  The sensitivity of those detectors have been studied assuming they are either capable of reconstructing full three-dimensional (3D) tracks~\cite{Green:2006cb,Morgan:2004ys} or only tracks projections on a particular detector-fixed plane (2D detector)~\cite{Green:2006cb,Morgan:2005sq}.  It is found that of order tens of events are required to reject isotropy of recoil angle distributions, and the number of required events are one order of magnitude larger for a detector that cannot measure the sense, defined as the absolute sign ($\vec x$ or $-\vec x$), of the recoil vector. However, it is still a daunting task to fully reconstruct sub-100 keV nuclear recoil tracks.

In this paper, we consider a detector that can detect the polar angle of the recoil track with respect to a fixed $z$ axis, while the azimuthal angle is not measured.  In the following, we refer to this as a polar detector. In general, since reconstruction of recoiling nucleus tracks is not required for a polar detector, the experimental realization can be more reliable and feasible.  One issue for a directional dark matter detector is whether or not it possesses sense recognition capability, where we call the former a head-tail detector and the latter an axial detector.  Considerable technological effort is required to provide head-tail detection capability.  Throughout the paper, the standard isothermal galactic halo model is used to model the WIMP velocity distribution.  In this case, a head-tail polar detector with $z$ axis aligned with the WIMP wind direction has the same performance as a full 3D detector, because in this orientation the azimuthal distribution is completely flat and provides no information at all.

One example of an axial polar detector is a stilbene crystal, which is an organic single crystal whose scintillation efficiency depends on the nuclear recoil direction relative to crystallographic axes~\cite{Sekiya:2003wf}.  Recently, a new technique of columnar recombination (CR) has been shown to be capable of realizing a working axial polar detector~\cite{Nygren:2013nda}, without sense detection capability.  The directional sensitivity in a CR detector comes from the dependence of electron-ion recombination level on the angle between recoil track and electric field.  Here the experimental signals are the scintillation light for the recombinations, and drift electrons from the surviving ionizations.

In most previous sensitivity studies such as those reported in Refs.~\cite{Green:2006cb,Billard:2009mf,Copi:2005ya,Morgan:2004ys,Mohlabeng:2015efa}, the recoil energy is integrated out and only distributions with respect to angular variables are studied.  When all the energies are integrated out, the shape of angular distribution does not depend on the WIMP mass~\cite{Alenazi:2007sy} and, thus, the distribution and the analysis are greatly simplified.  However, we note that in a real experiment, all relevant information are used in order to achieve the highest possible sensitivity.  Thus, both energy and directional information are used simultaneously in this work.  Previous studies that did use both the energy and angular information~\cite{Billard:2011zj,Billard:2014ewa}, treated the WIMP mass as a known parameter that was kept at a fixed value in the fit.  While this procedure is simpler to implement for producing plots of cross-section upper limits versus WIMP mass, it is problematic for a general directional sensitivity calculation. In this study, we address this issue by treating the WIMP mass as a nuisance parameter in a statistical framework that uses the well-established profile likelihood method.

We study signals from spin-independent WIMP-nucleus interactions, and assume zero background to provide benchmark results.  For nonzero background experiments, an observation of directionality still unambiguously leads to an observation of WIMPs, and the amount of directional sensitivity loss due to background is also studied.

In sections that follow, we will start by calculating the distribution of observables, and then find the best orientations for various detector configurations using the standard profile likelihood method. Later, the effects of background contamination are calculated for detectors in optimal orientations. Finally, we study the performance of detectors using xenon target with various energy thresholds and WIMP masses, and estimate the required detector exposure to see a three sigma directional signal.

\section{Distributions of observables}
The general double-differential derivative of the recoil rate per
unit target mass, with respect to nuclear recoil energy $E$ and
solid angle $\Omega$ in the nuclear recoil direction $\hat{q}$,
can be expressed as~\cite{Gondolo:2002np}:
\begin{equation}
  \frac{d^2R}{dEd\Omega} = \frac{n\sigma_0}{4\pi\mu^2}\mathcal{F}(E)
  \hat{f}(v_\mathrm{min},\hat{q}).
\end{equation}
Here $\mu=mM/(m+M)$ is the reduced mass of
the WIMP-nucleus system, where $m$ and $M$ are the WIMP and target nucleus mass, respectively. $\mathcal{F}$ is the form factor, $\sigma_0$ is
spin-independent WIMP-nucleus cross-section, and $n=\rho^0/m$
is the number density of the WIMP, where we use $\rho^0=0.3\ \mathrm{GeV/cm}^3$.
The minimum WIMP speed required for a recoil of energy $E$ is $v_\mathrm{min}$.
The Radon transform, \(\hat{f}(v_\mathrm{min},\hat{q}) \), 
of the WIMP velocity distribution \(f(\mathbf{v}) \) represents
the sum of the probability densities where the velocity projection on direction 
\(\hat{q}\) is equal to \(v_\mathrm{min}\):
\begin{equation}
  \hat{f}(v_\mathrm{min},\hat{q})
  = \int \delta(\mathbf{v}\cdot\hat{q}-v_\mathrm{min})f(\mathbf{v})d^3v.
\end{equation}
In the standard isothermal galactic halo model, 
the distribution of WIMP velocities $\mathbf{v}$ relative to the target
is Maxwellian, with an average value of $\mathbf{v_E}$:
\begin{equation}
  f(\mathbf{v})=\frac{1}{(\sqrt{\pi}v_0)^3}e^{-(\mathbf{v}-\mathbf{v_E})^2/v_0^2}.
\end{equation}
In this case the Radon transform is:
\begin{equation}
\hat{f}(v_\mathrm{min},\hat{q})
 = \frac{1}{\sqrt{\pi}v_0}e^{-(v_\mathrm{min}-\hat{q}\cdot\mathbf{v_E})^2/v_0^2}.
\end{equation}

The maximum recoil energy for a WIMP with velocity $v$ is
\(E_\mathrm{max}(v) = 2\mu^2v^2/M \), and by reversing the formula, we obtain
$v_\mathrm{min}=\sqrt{\frac{ME}{2\mu^2}}=\sqrt{\frac{E}{E_0}}v_0$, with the
definition $E_0=E_\mathrm{max}(v_0)$.  We also calculate the
standard total rate $R_0$ when $v_E=0$ to be
\(R_0 = n\sigma_0/M\cdot\int vf(\mathbf{v})|_{v_E=0}d^3v 
  = 2/\sqrt{\pi}\cdot n\sigma_0v_0/M\).  For a a single target-nucleus type,
the differential rate exhibits more physical meaning when
it is expressed using $E_0$ and $R_0$:
\begin{equation}
  \frac{d^2R}{dEd\Omega} = \frac{1}{4\pi}\frac{R_0}{E_0}\mathcal{F}(E)
  e^{-(v_\mathrm{min}-\hat{q}\cdot\mathbf{v_E})^2/v_0^2}.
\end{equation}

It is convenient to define two variables:
\begin{equation}
 x \equiv \frac{v_\mathrm{min}}{v_0}= \sqrt{\frac{E}{E_0}}; \quad
 x_E \equiv \frac{v_E}{v_0}.
\end{equation}
In a reference frame where the reference $z$ axis is parallel to 
the WIMP wind direction, or the Cygnus direction, the solid angle
dependence reduces to polar angle $\theta$ dependence, and
the differential rate can be expressed as:
\begin{equation}\label{eqn:dRdEdCos}
  \frac{d^2R}{dEd\cos\theta} = \frac{R_0}{2E_0}\mathcal{F}(E)
  e^{-(x_E\cos\theta-x)^2}.
\end{equation}

For simplicity, the peculiar velocity of the Sun and the Earth's orbital velocity about the Sun are ignored and the velocity of Earth $v_E$ is taken to be equal to the velocity of the Local Standard of Rest $\vec\Theta_\mathrm{LSR}$.  In this case the magnitude of $v_E$ and $v_0$ is same, which is usually given the value of 220 km/s.  Hence in this paper we use $x_E=1$ to represent the standard halo model.

It is feasible to build a detector that is oriented at a fixed direction in space, where the WIMP wind direction and the $z$ axis of detector form a fixed angle $\theta_0$.  In the frame of a detector system, the polar and azimuth angles of recoil $\theta_L$ and $\phi_L$ provide the directional information and the $\cos\theta$ term in Eq.~(\ref{eqn:dRdEdCos}) can be expressed in terms of $\theta_L$ and $\phi_L$ as
\(\cos\theta = \cos\theta_0\cos\theta_L + \sin\theta_0\sin\theta_L\cos\phi_L \).
Here the WIMP wind direction is assumed to lie in the $x-z$ plane, without any loss in generality.  The resulting differential rate in lab frame with respect to $E$, $\cos\theta_L$ and $\phi_L$ is then:
\begin{multline}
  \frac{d^3R}{dEd\cos\theta_L d\phi_L} = \\
\frac{R_0}{4\pi E_0}\mathcal{F}(E)
  e^{-(x_E\cos\theta_0\cos\theta_L+x_E\sin\theta_0\sin\theta_L\cos\phi_L-x)^2}.
\label{eqn:dRdEdCosLdPhiL}
\end{multline}
Here, a solid angle transform
\( \frac{1}{2\pi}\frac{d^2R}{dEd\cos\theta} \to
\frac{d^3R}{dEd\cos\theta_L d\phi_L} \) is applied
from Eq.~(\ref{eqn:dRdEdCos}).

For a detector where the only directional measurable is the polar angle
$\theta_L$, the azimuthal angle $\phi_L$ should be integrated out:
\begin{equation}
  \frac{d^2R}{dEd\cos\theta_L} = \int_0^{2\pi}\frac{d^3R}{dEd\cos\theta_L d\phi_L}.
\label{eqn:dRdEdCosL}
\end{equation}
For an axial detector where the sense of the recoil direction
cannot be distinguished, the distributions should be folded, as 
\( \frac{d^2R}{dEd|\cos\theta_L|d\phi_L} = 
 \frac{d^2R}{dEd\cos\theta_Ld\phi_L} + \frac{d^2R}{dEd(-\cos\theta_L)d(\phi_L+\pi)} \), and \( \frac{d^2R}{dEd|\cos\theta_L|} = 
\frac{d^2R}{dEd\cos\theta_L}+ \frac{d^2R}{dEd(-\cos\theta_L)} \).

\begin{figure}
\includegraphics[width=\columnwidth]{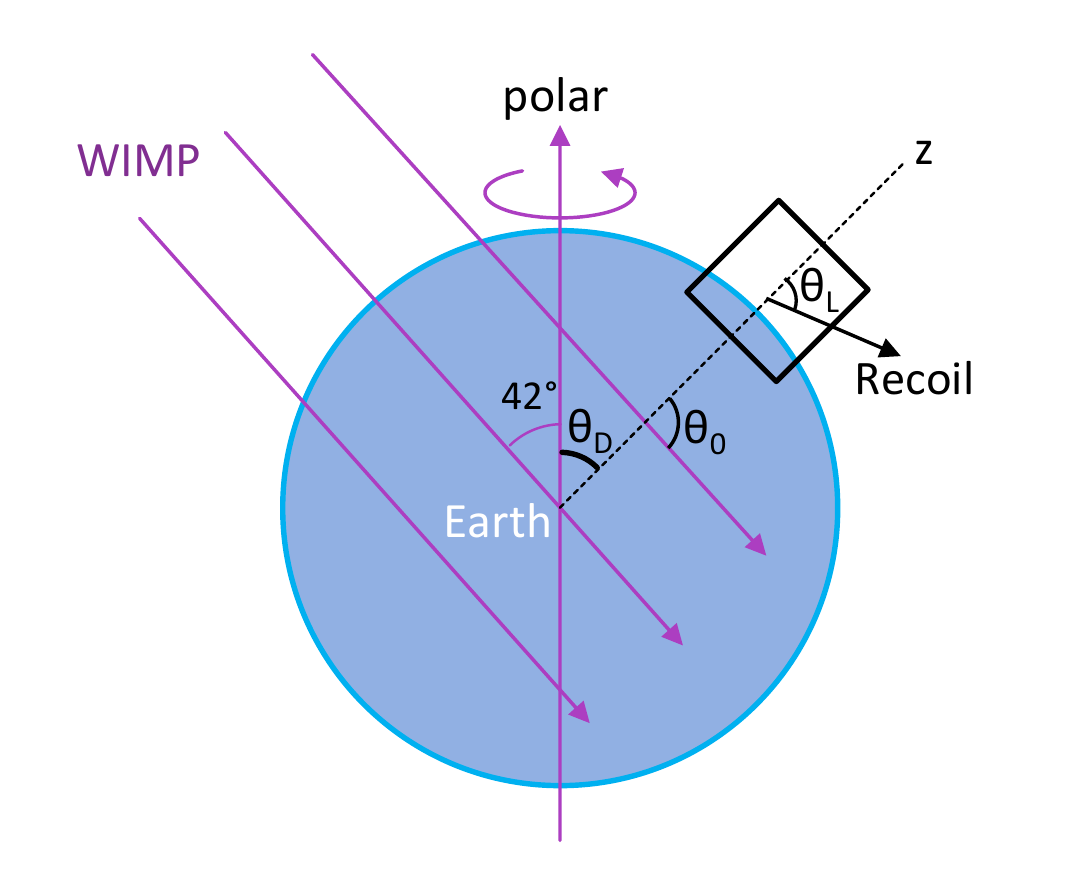}
\caption{Schematic plot of an Earth-fixed detector shown as a rectangular box. }
\label{fig:earth-DT}
\end{figure}

Now we consider a common case where a detector is fixed in the Earth's coordinate frame.  Since the Earth itself is rotating, the angle $\theta_0$ modulates with a period of one sidereal day.  The WIMP wind from Cygnus is directed to the Earth at a constant angle $\epsilon=42^\circ$ to the Earth's polar axis, as shown in Fig.~\ref{fig:earth-DT}.  If the $z$ axis of the detector is oriented relative to Earth's polar axis at a fixed angle $\theta_D$, then the dependence of the angle $\theta_0$ as a function of time $t$ in units of sidereal days is:
\begin{equation}
\cos\theta_0 = -\cos\epsilon\cos\theta_D - \sin\epsilon\sin\theta_D\cos(2\pi t).
\label{eqn:costheta0}
\end{equation}
Here $t$ is zero when the WIMP velocity lies in the plane made by the detector $z$ axis and Earth's polar axis.  Since $t$ distributes uniformly within 0 and 1 when all energies and angles are considered, the appropriate distribution for the Earth-fixed case, now in terms of ($E,\cos\theta_L,t$), $\frac{d^3R}{dEd\cos\theta_Ldt}$, is determined by replacing $\theta_0$ in Eq.~(\ref{eqn:dRdEdCosLdPhiL}) with Eq.~(\ref{eqn:costheta0}) and integrating out $\phi_L$.

Note that here, instead of using angular information only as in~\cite{Green:2006cb,Billard:2009mf,Copi:2005ya,Morgan:2004ys}, we use two and three dimensional distributions that include energy and sidereal time as additional variables.

\section{Statistical test for directional signal}
The most common method that is used in particle physics to determine
significance of an observation is the profile likelihood ratio
test statistic method.  It uses a hypothesis test
against a null or trivial hypothesis $H_0$, where the data is assumed
to correspond to the distribution of an alternative or interesting hypothesis
$H_1$.  Usually, a set of parameters of interest $\nu$ of $H_1$
are fixed to a specific value $\nu_0$ to obtain $H_0$.
A likelihood ratio is calculated as the maximum likelihood of
a null hypothesis divided by the maximum likelihood of an alternative
hypothesis, with nuisance parameters $\mathbf{\theta}$ floating:
\begin{equation}
\lambda(\nu_0) = \frac{L(\nu=\nu_0,\hat{\hat{\mathbf\theta}})}
      {L(\hat\nu,\hat{\mathbf\theta})}.
\label{eqn:lratio}
\end{equation}

The test statistic $q$ is defined as $-2\ln\lambda(\nu_0)$.
The $p$ value is the probability to have a discrepancy larger
than the observed one $q^\mathrm{obs}$, as
\begin{equation}
p = \int_{q^\mathrm{obs}}^\infty f(q|H_0)dq_0 .
\end{equation}
The smaller the $p$ value, the more credible that hypothesis $H_0$ is
not correct.  For example, a $p$ value of 0.00135 corresponds to a
$3\sigma$ signal significance.
In principle, the distribution \(f(q|H_0)\) needs to be
obtained using Monte Carlo methods with high statistics.
However, according to Wilk's theorem, we can assume $q$ follows a $\chi^2$
distribution~\cite{Cowan:2010js} for data sets in our study.
Then the $p$ value can be directly
calculated as the probability above $q^\mathrm{obs}$
for a $\chi^2$ distribution $P$ with degrees of freedom 
equal to the number of fixed parameters
in the numerator of Eq.~(\ref{eqn:lratio}):
$p=\int_{q^\mathrm{obs}}^\infty P(x)dx$.  The significance $Z$ is
related with $p$ via $Z=\Phi^{-1}(1-p)$, where $\Phi$ is
the cumulative function of the standard Gaussian.

The likelihood function $L$ used in this paper is the unbinned product of normalized probability density function (PDF) of observed quantities for all events.  The PDF is normalized over two or three dimensional variables, such as $(E,\cos\theta_L)$ or with the inclusion of $t$.  A nonzero $x_E$ indicates a finite average WIMP speed from the Cygnus direction. Thus, for studies in this paper, $x_E$ as $\nu$ is the sole parameter of interest, with the null hypothesis being $x_E=0$.  The WIMP mass, since it is unknown, is a nuisance parameter and floated in the fits for both the numerator and denominator of Eq.~(\ref{eqn:lratio}).  Unlike most of the studies such as~\cite{Billard:2014ewa}, where a fixed WIMP mass need to be inserted in order to get the sensitivity or limit, our method directly focuses on the directionality itself.

To test the sensitivity of a measurement, the median (not mean) value of the test statistic $q_\mathrm{med}$ of a large ensemble of hypothesis tests is used as a measurement quantity of sensitivity for the work reported here.  The ensemble size $N$ is 1000 for all of calculations discussed below.  The error of $q_\mathrm{med}$ is $\frac{1}{2\sqrt{N}f(q_\mathrm{med})}$, where $f(q)$ is the normalized probability density function that is obtained from the raw distribution of $q$ for the ensemble by a kernel density estimation method~\cite{Cranmer:2000du}.  Here, for each hypothesis test, a toy Monte Carlo data set with a certain number of events is generated and fitted to obtain the $q$ value for this data set.

With only one parameter of interest, the significance in units of $\sigma$ is calculated as the square root of the test statistic: $Z=\sqrt{q}$.  So a $Z=3\sigma$ significance corresponds to $q=9$.  In all the studies of this work, $q_\mathrm{med}$ is found to be proportional to the total event number, as will be shown in Fig.~\ref{fig:nE_q-n10}.  So we use a linear proportional function to fit $q_\mathrm{med}$, and find the number of events corresponding to $q_\mathrm{med}=9$ as the required size for a $3\sigma$ discovery.  Because of the linearity, the sensitivity can also be scaled proportionally to the size of statistics when needed.

\begin{figure}[t]
\includegraphics[width=\columnwidth]{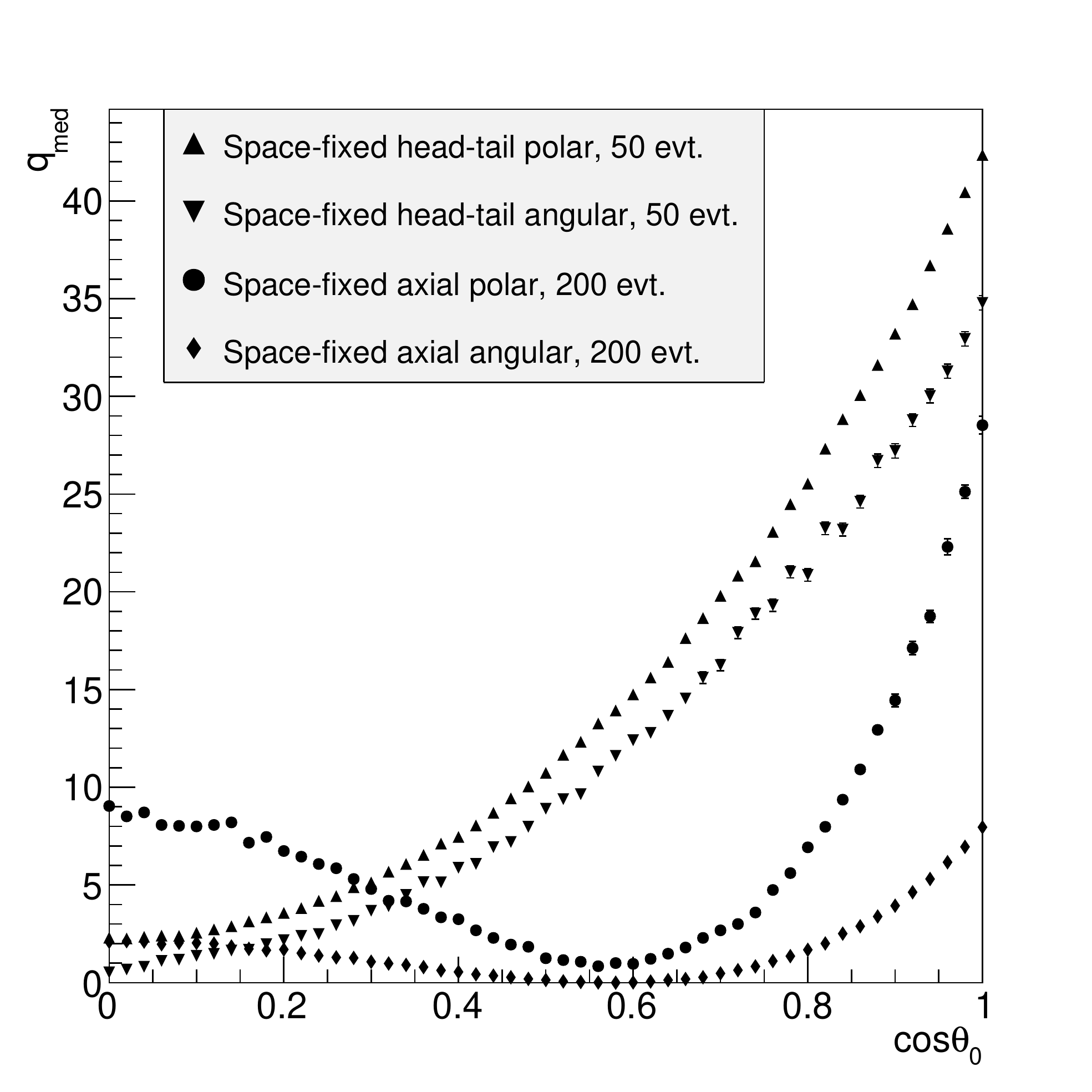}
\caption{Median $q$ values on solid markers as a function of cosine of the angle between the detector $z$ axis and the WIMP wind direction for general space-fixed detectors.  Head-tail polar and angular only detectors correspond to up and down triangles.  Axial polar and angular only detectors correspond to circles and diamonds.   The event number for head-tail configuration is 200, different from 50 events for axial configuration, in order to fit in a single plot. }
\label{fig:q_vc-SF4}
\end{figure}

\begin{figure}[t]
\includegraphics[width=\columnwidth]{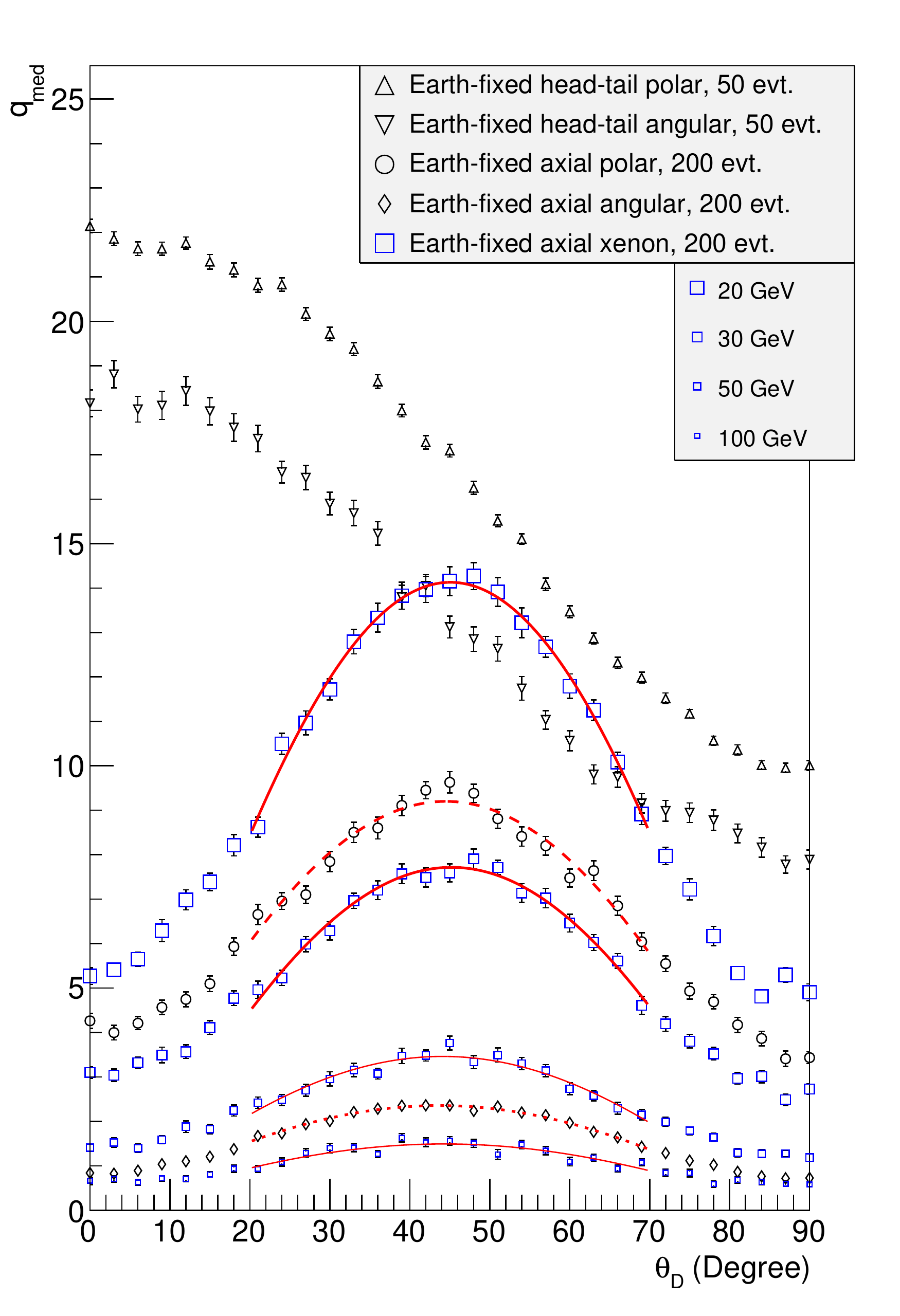}
\caption{Median $q$ values on open markers as a function of angle $\theta_D$ between detector $z$ axis and Earth's pole, for Earth-fixed detectors.  General head-tail polar and angular only detectors with 50 events correspond to up and down triangles.  General axial polar and angular only detectors with 200 events correspond to circles and diamonds.   For a standard axial xenon detector of 3 keV threshold with 200 observed events, four graphs for WIMP masses of 20, 30, 50 and 100 GeV are shown as blue squares with decreasing sizes. These graphs are fitted using an empirical function $q_\mathrm{med}= Q\cos(\alpha(\theta_D-\theta_D^0))$ that are superimposed as red solid curves for a standard axial xenon detector, and as red dashed and dotted curves for general axial polar and angular only detectors.}
\label{fig:q_vc-RT5}
\end{figure}

\begin{figure}
\includegraphics[width=\columnwidth]{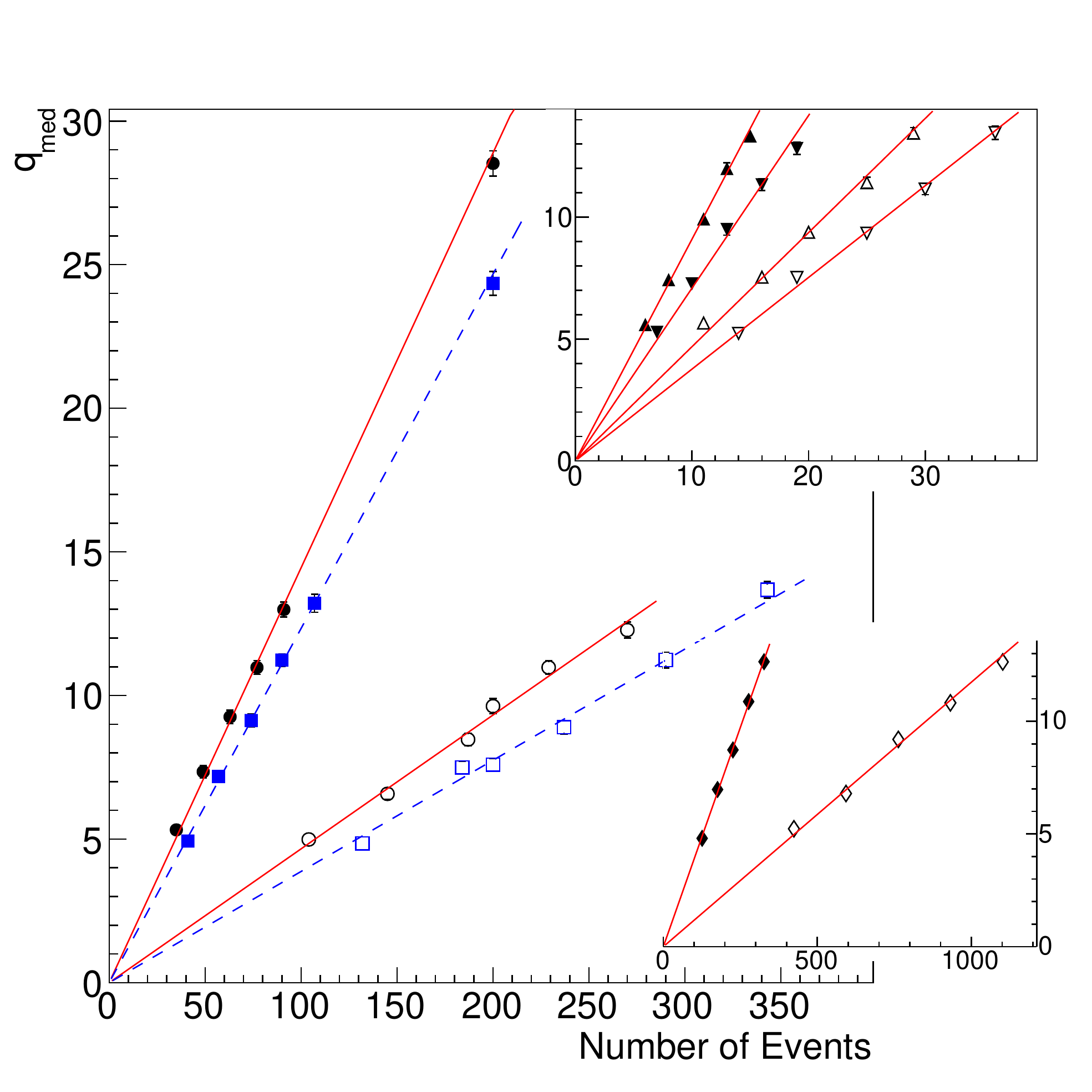}
\caption{Median $q$ values versus numbers of observed events for all detector types in their optimal orientations.  Solid squares represent a standard space-fixed axial xenon detector.  Other symbol markers are explained in Figs.~\ref{fig:q_vc-SF4} and \ref{fig:q_vc-RT5}.  The fitted linearly proportional lines are superimposed.  Blue dashed lines represent axial xenon detectors configured at 30 GeV WIMP and 3 keV threshold.  }
\label{fig:nE_q-n10}
\end{figure}

\section{General Polar Detector} \label{sec:gen}
To study the general performance of a polar detector, a general detector with zero energy threshold and unit form factor is considered.  In the absence of the form factor term in Eq.~(\ref{eqn:dRdEdCosLdPhiL}),  the energy dependence of the rate is expressed through the ratio $E/E_0$, i.e., in units of $E_0$.  Since the WIMP mass dependence of the rate function only enters through $E_0$, a change of WIMP mass just changes $E_0$, the overall scale of the energy $E$.  Thus the sensitivity for the general polar detector has no dependence on the WIMP mass.  In Secs.~\ref{sec:gen}--\ref{sec:bkg}, general results applicable to all WIMP masses are shown.

Figure~\ref{fig:q_vc-SF4} shows the median value of $q$ for space-fixed detectors, with head-tail and axial configurations, for different detector orientation angles.  We can see that the optimum orientation for a space-fixed case is always at $\cos\theta_0=1$, where the detector's $z$ axis points toward the Cygnus.  This is expected since in this case the detector can gain maximum information from the measured $\cos\theta_L$.  Nontrivial curves for axial detectors are discovered with a minimum at around $\cos\theta_0 = 0.56$, where the distribution of $|\cos\theta_L|$ is close to flat due to the folding of positive and negative $\cos\theta_L$ values. 

In the Earth-fixed case, the quantity $q_\mathrm{med}$ is averaged over many directions, and depends on the orientation $\theta_D$.  Figure~\ref{fig:q_vc-RT5} shows the dependence of $q_\mathrm{med}$ on the angle $\theta_D$.  A general head-tail detector is shown to have the most sensitivity when the $z$ axis is oriented parallel to the Earth's polar axis ($\theta_D=0^\circ$).   Interestingly, the optimal $\theta_D$ value for a general axial detector is near $\theta_D=45^\circ$.  A fit using an empirical function $q_\mathrm{med}= Q\cos(\alpha(\theta_D-\theta_D^0))$ gives the optimum value $\theta_D^0=44.48 \pm 0.33$, with the fitted curve shown in Fig.~\ref{fig:q_vc-RT5} as a dashed red line.  The reason for $\theta_D^0$ being slightly larger than 42 degree, is that it is more advantageous to lie inside than outside for the WIMP wind vector, with respect to the cone made by the detector's $z$ axis following the Earth's self-rotation.   Note that this value $\theta_D^0$  will change if the angle of the WIMP wind to the Earth's pole is a value different than 42 degree.

Results for the required number of events for $3\sigma$ discovery for four detector types, i.e., $\cos\theta_0=1$ for space-fixed detectors, and $\theta_D=45(0)^\circ$ for Earth-fixed axial (head-tail) detectors, are displayed in row one of Table~\ref{tab:ensXE_nEq9_n12}.  Of order 10 to 20 event are required for head-tail detectors.  These results show that a change from a space- to Earth-fixed basis worsens the sensitivity by factors of $3.1$ and $1.9$ for axial and head-tail configurations.  For an Earth-fixed detector, we conclude that the loss of head-tail capability causes the sensitivity to be a factor of 10 worse, and the axial one is less efficient than a full 3D one (column 2)  by about a factor of 19.  These numbers can be compared to the 1D readout in Ref.~\cite{Billard:2014ewa}, which gives values of 8 and $\sim 10$ for the two factors, by measuring the signal yield with a constrained background.

As discussed in the Introduction, the optimal space-fixed head-tail detector behaves the same as a full 3D detector with its $z$ axis aligned along the WIMP wind.  By dividing the two numbers for the space-fixed case, we expect that an axial detector would have the same sensitivity as a full 3D detector, provided it accumulated $6.3$ times as many events.  This is a much better conclusion than that in Ref.~\cite{Billard:2014ewa},  which did not study a space-fixed detector.

\begin{table*} 
\caption{The required number of point interactions for $3\sigma$ discovery for all detector types.  Numbers are obtained from the abscissa value corresponding to $q_\mathrm{med}=9$ level in Fig.~\ref{fig:nE_q-n10} for the fitted straight lines.  For xenon detector types, the numbers are calculated by applying the ratio $N_\mathrm{pint}/N_\mathrm{obs}$ from the required number of observed events.  The indicated errors are purely statistical from finite ensemble size of 1000.}
\label{tab:ensXE_nEq9_n12}
\begin{tabularx}{0.9\textwidth}{l|>{\centering}X|>{\centering}X|>{\centering}X|>{\centering\arraybackslash}X}
\hline \hline
\multirow{2}{*}{Detector type} &
\multicolumn{2}{c|}{space-fixed} &  \multicolumn{2}{c}{Earth-fixed} \\
 \cline{2-5}
      & Axial & Head-tail & Axial & Head-tail \\
\hline
General  & $62.30\pm 0.56$ & $ 9.90\pm 0.09$ & $193.2\pm 2.1$ & $19.22\pm0.19$\\
Angular Only & $231.4\pm 2.7$ & $12.70 \pm 0.13$ & $767.0\pm 9.2$ & $23.91\pm 0.25$ \\
Xenon, 30 GeV WIMP, 3 keV threshold   & $168.1 \pm 1.6$ & & $535.6 \pm 5.8$  &  \\
Xenon, 30 GeV WIMP, 10 keV threshold  & $218.5 \pm 2.0$ & & $702.7 \pm 7.6$  &  \\
Xenon, 50 GeV WIMP, 30 keV threshold  & $1318 \pm 12$ & & $4087  \pm 44$  &  \\
\hline \hline
\end{tabularx}
\end{table*}

\section{Angular only detector} \label{sec:ang}
Most previous estimates of the sensitivity of a directional dark matter detector have only considered angular information only.  To study detectors with polar angle information only, the rate as a function of $\cos\theta_L$ is obtained from Eq.~(\ref{eqn:dRdEdCosL}) by integrating out the energy part:
\begin{equation}
  \frac{dR}{d\cos\theta_L} = \int_0^\infty dE\frac{d^2R}{dEd\cos\theta_L} .
\end{equation}

The same statistical procedure is performed as for the general polar detector case, for space- or Earth-fixed, axial or head-tail configurations.  The $q_\mathrm{med}$ dependence on detector orientation is shown in Figs.~\ref{fig:q_vc-SF4} and \ref{fig:q_vc-RT5}.  In all four detector configurations, the shape of the $\cos\theta_0$ and $\theta_D$ dependence is similar to that for the general polar detector case.  For a space-fixed axial detector, the dependence on $\cos\theta_0$ is also not monotonic.  In the Earth-fixed case, the red dotted curve in Fig.~\ref{fig:q_vc-RT5} has a maximum at $\theta_D=43.68\pm 0.28^\circ$.  Results for the optimal value, $\cos\theta_0=1$ for space-fixed, $\theta_D=45^\circ$ and $0^\circ$ for Earth-fixed axial and head-tail detectors,  are reported in row 2 of Table~\ref{tab:ensXE_nEq9_n12}.  For head-tail detectors, of order of 10 (20) events are required to see directionality in space (Earth)-fixed basis, similar to results from previous studies~\cite{Morgan:2004ys,Morgan:2005sq,Copi:2005ya} for 3D and 2D detectors.  The sensitivity worsens by factors of $3.3$ and $1.9$ for axial and head-tail configurations, respectively, when changing a space-fixed to an Earth-fixed basis, similar to the general polar detector case.

By comparing the event numbers in columns 1 and 3 of Table~\ref{tab:ensXE_nEq9_n12}, i.e., for the axial configurations of general and angular only detectors, we can see that $3.7$ and $4.0$ larger statistical samples are needed when the energy information is not used.    However, for head-tail detectors (columns 2 and 4), only $1.3$ and $1.2$ times the statistics are required.  This is because in the absence of the powerful head-tail information, which directly relates to directionality, we need to rely on the correlation between recoil energy and direction to improve the directional sensitivity.  Thus, the importance of combining energy and angular observables is established, at least for axial detectors.  We note that the performance may be improved over that reported in Refs.~\cite{Morgan:2005sq,Copi:2005ya}, if the observed energy is simultaneously used.

\begin{figure}
\includegraphics[width=\columnwidth]{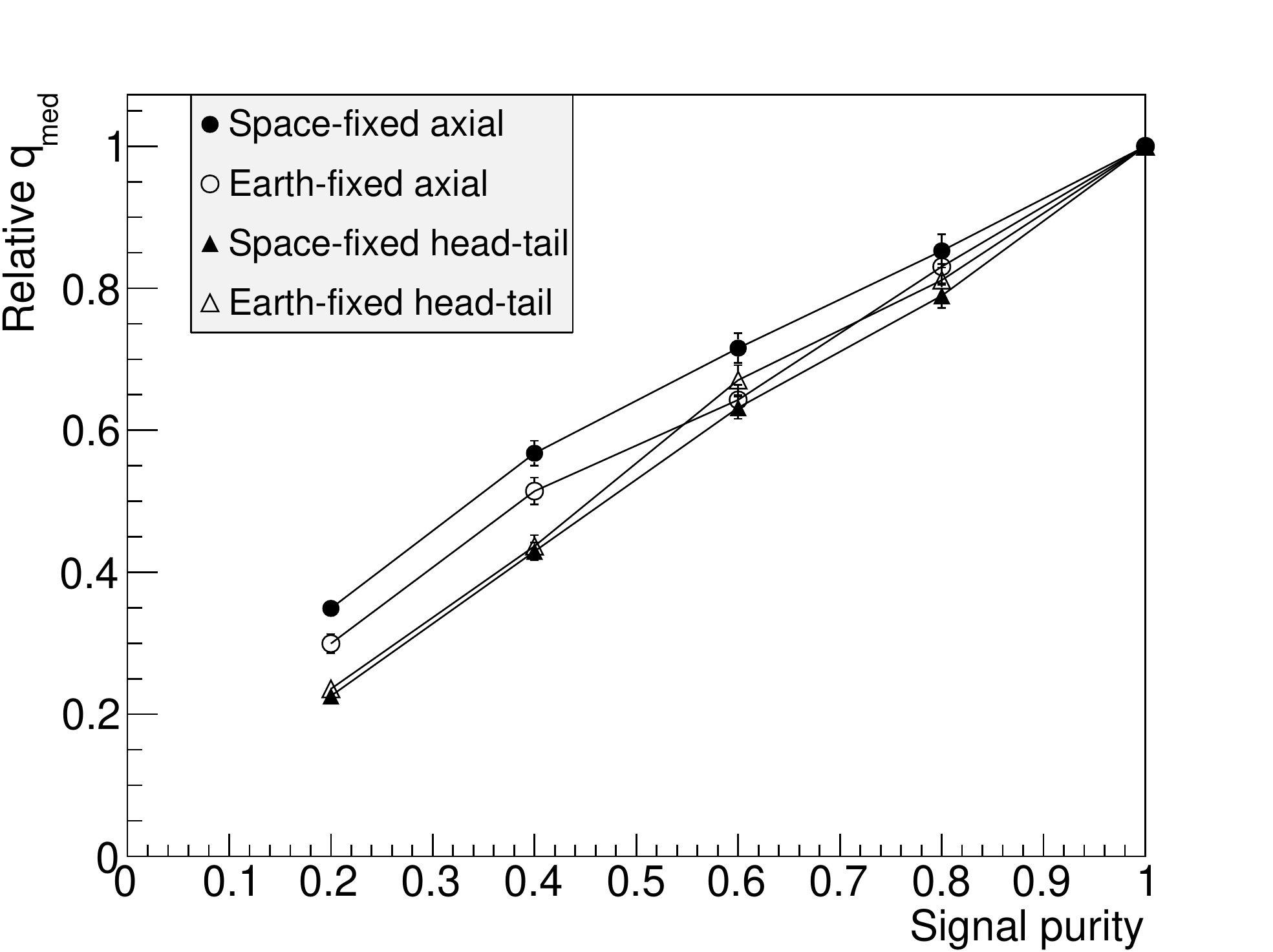}
\caption{Sensitivity ratio over that for pure signal case, as a function of signal purity defined as the ratio of signal component over all events in the data set.  For each detector type, data points are normalized to a constant number of signal events.  Space- and Earth-fixed polar detectors under axial and head-tail configurations are shown when placed in each one's optimal orientation.}
\label{fig:q_fs-SB4}
\end{figure}

\section{Background effect}  \label{sec:bkg}
In nondirectional dark matter searches, the most dangerous backgrounds are those with the same energy-dependent shape as the signal, since in those cases one cannot distinguish the background by the energy distribution.  In this study, we use the directional observables to distinguish the signal, while the directional information is diluted by the background contribution.

Consider a background that is distributed the same way as the WIMP signal, while not requiring the relative speed between the earth and the dark matter halo.  By setting $x_E=0$ and a unit form factor in Eq.~(\ref{eqn:dRdEdCosLdPhiL}), the background rate $R_\mathrm{BG}$ is isotropically distributed with a simple exponential decaying energy dependence: $\frac{d^2R_\mathrm{BG}}{dEd\cos \theta_L} = \frac{R_0}{2E_0}e^{-E/E_0}$.

In a real experiment, usually we do not fully understand all the sources of background.  Here we show that this is not a problem in our method.  Using the same PDF [Eq.~(\ref{eqn:dRdEdCosL})] to fit data, regardless of the how the data is generated or originated as would be in a real experiment, a nonzero $x_E$ still unambiguously indicates a directional signal.  We use the same statistical procedure, and only change the generation of the signal-only data set to a mixture of signal and background.  The parameter of interest is still $x_E$.  Our method avoids the need for an external constraint to be imposed on the background rate, as done, for example, in Ref.~\cite{Billard:2014ewa} to obtain a better control of the signal component.

The sensitivity dependence on the signal purity is shown in Fig.~\ref{fig:q_fs-SB4}, where the signal component yield is a constant for each detector type.  All polar detectors are aligned in their optimal orientations.  The $x$-axis is the signal purity, defined as the ratio of the input signal number over total event number in the generation.  In Fig.~\ref{fig:q_fs-SB4}, the sensitivity level can be seen to be almost linear with the signal purity.  As the signal purity goes lower, the relative sensitivity falls by a lesser amount, especially for the space-fixed axial case.  Thus we conclude the polar detector has a guaranteed performance level in the presence of background.  In particular, the required number of events for purity 0.4 only increases by about a factor of 2 compared to the pure signal case.  This can be compared to the 1D case of Fig.~2 in Ref.~\cite{Billard:2014ewa}, in which case the degradation of sensitivity is significantly worse as the signal purity decreases.

Note that, other types of backgrounds, such those with a flat energy dependence, are easier to handle because the difference in the energy distribution gives extra information to distinguish them from the real signal.

\section{Standard Xenon Detector}
Now we consider a detector with a specific material and a minimum energy threshold.  Since the proposed columnar recombination detector uses gaseous xenon as target, a xenon Helm form factor~\cite{Lewin:1995rx} $\mathcal{F}(E)$ is computed and inserted into Eq.~(\ref{eqn:dRdEdCosLdPhiL}).   For the minimum energy threshold, the LUX experiment~\cite{Akerib:2013tjd} provides a good reference, where the trigger efficiency reaches 50\% for 4 keV nuclear recoil events.   In gaseous xenon, the commonly used reflective material polytetrafluoroethylene (PTFE) has a low refractive coefficient of about 60\% for the 175 nm scintillation wavelength of pure xenon, compared to more than 90\%~\cite{Yamashita:2004} in liquid xenon.  An experimental prototype~\cite{Renner:2014mha} using pure xenon has put a high threshold of 30 keV, mainly due to a low light collection efficiency of 3\%.

A small amount of trimethylamine (TMA) mixture in the xenon gas, can convert the primary excitation to ionization of TMA molecules through the Penning transfer process~\cite{Cebrian:2012sp}, and increases the CR effect since more ionizations will participate in the recombination process~\cite{Nygren:2013nda}.  At the same time,  the scintillation light is wavelength-shifted to the TMA emission band centered at near-UV (300 nm), where the reflectance of PTFE increases to 90\%~\cite{Silva:2009ip}.  Thus, it should be possible to achieve an energy threshold much lower than 30 keV.

In this study, we consider three minimum energy threshold values of 3, 10 and 30 keV.  Because of the rapidly falling shape of the form factor, the event rate will be suppressed by a factor more than $10^{-6}$ for recoil energies greater than 100 keV. Thus, the maximum of the detection energy range is chosen to be 100 keV.

This standard xenon detector can be viewed as a columnar recombination detector with perfect resolution.  So we only study the axial configuration.  First, in the Earth-fixed case, the sensitivity dependence on $\theta_D$ is studied for four typical WIMP mass values of 20, 30, 50 and 100 GeV.  Figure~\ref{fig:q_vc-RT5} shows the results for a 3 keV threshold, with fitted curves superimposed.  The fitted optimal values for $\theta_D $ are $45.08\pm 0.22, 45.21\pm 0.30, 44.12\pm 0.49, 44.30\pm 0.79$ degree for 20, 30, 50 and 100 GeV WIMP mass.  These are all nearly equal; a weighted average value is  $44.97\pm 0.17$.  In the following, the optimal directions $\theta_D=45^\circ$ and $\cos\theta_0=1$ are used for Earth- and space-fixed standard xenon detectors.

Because of the detection threshold, the required number of observed events for detectors with different thresholds cannot be directly translated to detector performance, since in this case the total rate depends on the threshold.  The number of observed events, $N_\mathrm{obs}$, is the integral of the differential rate within the detection range including the form factor for a given total target mass $M_T$ and measurement time $T$:
\begin{equation}
 N_\mathrm{obs}= RM_TT=\int_\mathrm{det. range} \frac{d^2R}{dEd\Omega}M_TT. 
\end{equation}
To make fair comparisons of the sensitivities, we introduce a quantity called the number of point interactions, $N_\mathrm{pint}$, which is the number of WIMP-nucleus interactions with no threshold and no form factor effects:
\begin{equation}
\begin{split}
 N_\mathrm{pint}&= \int_0^\infty dE\int d\Omega \frac{d^2R}{dEd\Omega}|_{\mathcal{F}(E)=1}\\
 &=1.304R_0M_TT=2.608/\sqrt{\pi}\cdot n\sigma_0v_0M_TT/M. 
\end{split}
\end{equation}
 The amount of required $N_\mathrm{pint}$ directly relates to detector performance.  For general detectors discussed in previous sections, $N_\mathrm{obs}= N_\mathrm{pint}$.  For a single target, the spin-independent WIMP-nucleus cross-section $\sigma_0$ is related to the WIMP-nucleon cross-section $\sigma_n$ as $ \sigma_0 = \mu/\mu_{W-n}A^2\sigma_n$, where $\mu_{W-n}$ is the WIMP-nucleon reduced mass.  For xenon, $N_\mathrm{pint}= 1.70\times 10^{46}\sigma_n M_{T} T/m \cdot \mathrm{GeV\cdot (cm^{2}\cdot kg\cdot year)^{-1}}$.  

\begin{figure}
\includegraphics[width=\columnwidth]{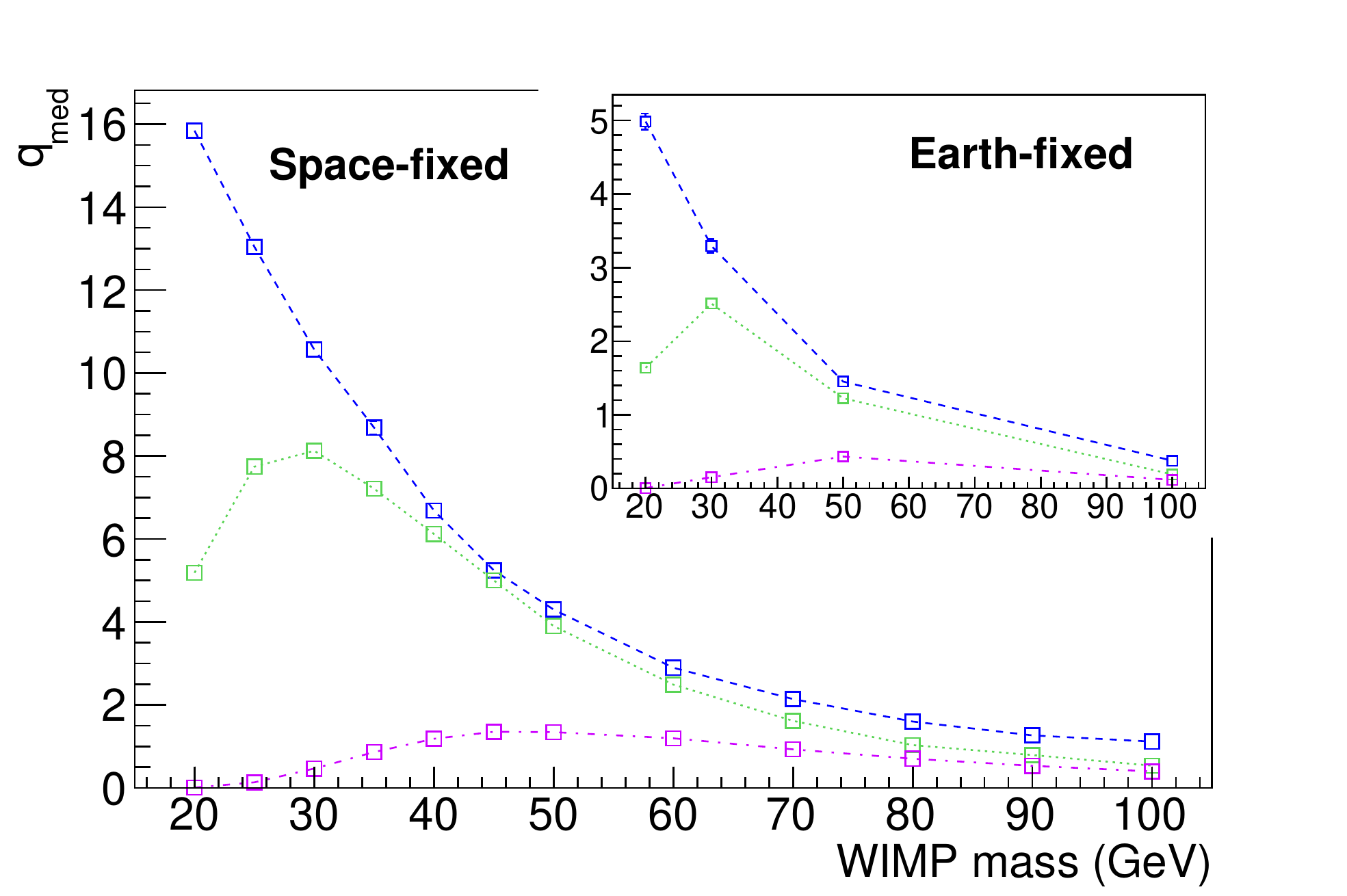}
\caption{Median $q$ values as a function of WIMP mass for axial xenon detectors, normalized to 200 events of point interaction.  From the top to bottom, the dashed, dotted and dot-dashed lines corresponds to the 3, 10 and 30 keV energy threshold.  The best orientations on space-fixed ($\cos\theta_0=1$, main plot) and Earth-fixed ($\theta_D=45^\circ$, inset plot) basis are used.  } 
\label{fig:q_mW-xffc2-MW2}
\end{figure}

There will be some WIMP-mass dependence of the sensitivity since the form factor and the threshold depends on absolute energy values that do not scale with $E_0$.   The WIMP mass is still a floating nuisance parameter in the fitting, while being fixed at various input values in event generation.  Figure~\ref{fig:q_mW-xffc2-MW2} shows the raw sensitivity as a function of WIMP mass, for space- and Earth-fixed detectors at their optimal orientations.  Each point in the plot is normalized to $N_\mathrm{pint}=200$, so that the sensitivities can be compared for the same detector exposure.

In general, the higher the threshold, the lower the sensitivity. The decrease of the sensitivity for low WIMP masses is due to the effect of the threshold that eliminates a higher fraction of events the lower the WIMP mass.  The sensitivity decrease for high WIMP masses is due to the form factor, which suppresses the high energy recoils.  The most sensitive WIMP-mass value increases as the energy threshold increases, and is around 30 GeV for a 10 keV threshold and 50 GeV for a 30 keV threshold. The inset of Fig.~\ref{fig:q_mW-xffc2-MW2} shows that the performance of Earth-fixed detectors follows the same behavior as space-fixed case with respect to WIMP-mass and energy-threshold dependence, except for an overall sensitivity that is lower by a factor of about three, which is consistent with the general study in Sec.~\ref{sec:gen}.
 
We can also compare the decrease in sensitivity to the decrease in the observed event rate due to the energy threshold cut.  For example, for a 30 GeV WIMP, an energy threshold change from 3 keV to 10 keV, decreases the sensitivity  by only 23\%, compared to the 71\% decrease in the total number of observed events.  This is due to the fact that most directional information comes from the high recoil energy events, so that a cut on low energy events has minor impact.   For a detector with high energy threshold of 30 keV, a WIMP mass in the intermediate range around 50 GeV is most promising, with the sensitivity that is 3.2 times below that of a 3 keV-threshold detector for the same range.

Last three rows of Table~\ref{tab:ensXE_nEq9_n12} show the required point interaction numbers for a $3\sigma$ discovery under several WIMP-mass and energy-threshold combinations.  In a typical case of a 30 GeV WIMP mass and a 10 keV energy threshold, the numbers are $219$ and $703$, for space- and Earth-fixed xenon polar detectors.  These correspond to $770$ and $2480$ kg$\cdot$year exposures for a $5\times 10^{-46}\ \mathrm{cm}^2$ spin-independent WIMP-nucleon cross-section.

\section{Discussion}
A columnar recombination detector using high pressure xenon-TMA gas mixture can measure the angle between the nuclear recoil track and electric drift field~\cite{Nygren:2013nda,Gehman:2013mra}, without sense recognition capability.  Thus it is an axial polar detector.   The columnar effect benefits from a high Penning efficiency as the percentage of the xenon excitations that fall back to free electrons by ionizing the TMA molecules.  For electron and gamma energy deposit, the Penning efficiency is estimated to be around 10\% and 20\% in Ref.~\cite{Gonzalez-Diaz:2015nba} and Ref.~\cite{Ruiz-Choliz:2015daa}, and also indirectly measured to be 10\%-15\% in high pressure xenon gas with TMA mixture~\cite{Nakajima:2015cva}.  In addition, the primary scintillation from xenon is observed to be absorbed by the TMA mixture~\cite{Nakajima:2015cva}.  Nevertheless, the columnar recombination effect for the ionization channel has been seen for the alpha particle events as shown in Fig.~5 of Ref.~\cite{Herrera2014}, where, as a first approximation, the collected charge is a linear function of $\cos^2\theta_L$.  Further studies are needed to obtain a high Penning efficiency for the high initial ionization density produced by nuclear recoils.

To date, no experimental results on the properties of nuclear recoils in a columnar recombination material is available.   Once a reliable detector response is available, a detector resolution can be straightforwardly included in our model to obtain the sensitivity.  Despite this limitation, we have tested a naive resolution smearing method for nuclear recoils based on the following considerations: (a) the recombination fraction of the electron-ion pairs ranges from 0.8 for recoils that are parallel($\theta_L=0^\circ$) to 0.4 for those that are perpendicular ($\theta_L=90^\circ$) to the field direction; (b) an ionization work function of 76 eV for the electron-ion pairs; and (c) 10\% and 50\% efficiency for the recombination (scintillation) and ionization channels with Gaussian resolutions calculated from a Fano factor~\cite{Fano:1947zz} of 0.14~\cite{Alvarez:2012kua} for xenon gas.  The resulting sensitivity is only worsened by 7\% relative to the zero resolution case.   This can be understood as that the directionality is derived from an overall anisotropic phenomenon and not from a localized spot.  Reference~\cite{Mohlabeng:2015efa} also noted that variations in angular resolution do not make much difference in the dark matter detection sensitivity.  Thus we do not expect a significant change to our sensitivity results by detector resolution effects.

\section{Conclusion}
The performance of a directional dark matter detector with polar angle detection is studied for various configurations.   A WIMP-mass independent method is used to obtain the sensitivity of a general detector.  In addition, a detector with xenon as target material is studied.   We infer that:
\begin{itemize}
\item  Both axial and head-tail polar detectors have the highest sensitivity when the $z$ axis is aligned with WIMP wind.  However, the dependence of sensitivity to detector orientation is not monotonic for an axial detector.  To obtain optimal performance when rotating with the Earth, the $z$ axis should be oriented at 45 degree to Earth's pole for an axial detector, while it should be aligned with Earth's polar axis for a head-tail detector.
\item  A head-tail polar detector can detect directionality with of order 10 or 20 events on a space- or Earth-fixed basis.  In the absence of sense detection capability, an order of magnitude more statistics is needed.
\item  Without using energy information simultaneously, the required statistics would be a factor of $3.7\ (4.0)$ times higher for a space (Earth)-fixed axial detector.  This conclusion will be useful for detector types in which partial directional information is available, such as a 2D planar detector.
\item  A general axial polar detector with $6.3$ times the statistics has the same performance as a general full 3D tracking detector.  However, in experimental practice, the target mass for a full 3D detector is limited because of diffusion effects, and accomplishing millimeter tracking is extremely challenging. On the other hand, a detector with polar angle sensitivity without head-tail discrimination requirement can use straightforward experimental techniques.  In addition, it can be made in large volume with high-density gas, with a target mass that can be orders of magnitude larger than a conventional full 3D detector. We conclude that it is of great advantage to explore the directional dark matter detection technique using a senseless polar angle detection apparatus.
\item A space-fixed detector is generally found to be 3 and 2 times more sensitive than an Earth-fixed detector, for axial and head-tail configurations.  This ratio is an important factor when comparing the additional cost for a space fixed detector, since it has to rotate all the time with respect to the Earth.  In a space-fixed detector, to distinguish between the WIMP signal from the galaxy coordinate and possible anisotropic background originating from the detector frame, manually reversing the detector $z$ axis direction for half of the measurement period should be useful.
\item  In the presence of contamination by an isotropic background that mimics the energy shape of the signal, the decrease of directional sensitivity occurs at a rate that is less than the decrease of signal purity.  Thus a polar detector is robust against background.   The space-fixed axial configuration is least affected by such backgrounds.
\item  For an axial xenon polar detector, the decrease of sensitivity due to the effect of the energy threshold is much smaller than the decrease of the total event rate.   A $770$ or $2480$ kg$\cdot$year exposure can reach a $3\sigma$ directional signal on a space- or Earth-fixed basis with 10 keV energy threshold,  for a $5\times 10^{-46}\ \mathrm{cm}^2$ WIMP-nucleon cross-section and a 30 GeV WIMP mass.

\end{itemize}

\begin{acknowledgments}
The author would like to thank Adam Para, Jonghee Yoo, Yeongduk Kim and Stephen Olsen for useful discussions.  This work was supported by the Institute for Basic Science (Korea) under project code IBS-R016-D1.
\end{acknowledgments}

\end{document}